\documentclass[useAMS,usenatbib,fleqn]{mn2e}
\usepackage{graphicx}
\usepackage{amssymb}
\usepackage{times}
\usepackage{enumerate}

\def\aj{AJ}%
\def\araa{ARA\&A}%
\def\apj{ApJ}%
\def\apjl{ApJ}%
\def\apjs{ApJS}%
\def\aap{A\&A}%
\def\aapr{A\&A~Rev.}%
\def\mnras{MNRAS}%
\def\prd{Phys.~Rev.~D}%
\def\pasp{PASP}%
\def\nat{Nature}%
\title[Direct formation of MSPs from delayed AIC]{
Direct formation of millisecond pulsars from rotationally delayed accretion-induced collapse of massive white dwarfs
} 

\author[Freire \& Tauris]
{Paulo~C.~C.~Freire$^{1}$\thanks{E-mails: pfreire@mpifr-bonn.mpg.de / tauris@astro.uni-bonn.de},
Thomas M. Tauris$^{2, 1}$\footnotemark[1]
\\
$^{1}$ Max-Planck-Institut f\"ur Radioastronomie, Auf dem H\"ugel 69, D-53121 Bonn, Germany\\
$^{2}$ Argelander-Institut f\"ur Astronomie, Universit\"at Bonn, Auf dem H\"ugel 71, D-53121 Bonn, Germany.
}
\begin{document}

\maketitle

\begin{abstract} 
    Millisecond pulsars (MSPs) are believed to be old neutron stars, formed via type~Ib/c core-collapse supernovae, 
    which have subsequently been spun up to high rotation rates via accretion from a companion star
    in a highly circularised low-mass X-ray binary. The recent discoveries of Galactic field binary MSPs in eccentric orbits, 
    and mass functions compatible with that expected for helium white dwarf companions, PSR~J2234+06 and PSR~J1946+3417, therefore challenge this picture.ms.tex
    Here we present a hypothesis for producing this new class of systems, where the MSPs are formed
    directly from a rotationally-delayed accretion-induced collapse of a super-Chandrasekhar mass white dwarf.
    We compute the orbital properties of the MSPs formed in such events and demonstrate that 
    our hypothesis can reproduce the observed eccentricities, masses and orbital periods of the white dwarfs, 
    as well as forecasting the pulsar masses and velocities. 
    Finally, we compare this hypothesis to a triple star scenario.
\end{abstract}

\begin{keywords}
stars: neutron --- white dwarfs --- stars: rotation --- 
X-rays: binaries --- supernovae: general --- pulsars: general 
\end{keywords}


\section{Introduction}\label{sec:intro}
Almost since the discovery of PSR~B1937+21, the first millisecond pulsar \citep[MSP,][]{bkh+82},
it has been suggested that these objects are old neutron stars spun up to high spin frequencies
of several hundred Hz via accretion of mass and angular momentum from a companion star \citep{acrs82,bv91}.
In this so-called recycling phase, the system is first observable as a low-mass X-ray binary \citep[LMXB, e.g.][]{bcc+97},
later as an accreting X-ray MSP \citep{wv98}, or even as an MSP in the transition phase between
an accretion powered MSP and a rotation powered radio MSP \citep{asr+09,pfb+13,tau12}.

An inevitable consequence of a long phase ($10^8-10^9\;{\rm yr}$) of recycling in an LMXB, where tidal forces operate, is that it should leave a fossil record of a
highly circular system \citep{pk94}. And indeed, until recently, {\it all} of the more than 100 observed, 
fully recycled MSPs (here defined as pulsars with spin periods less than 20~ms),  
in binaries with helium white dwarf (He~WD) companions and located outside of globular clusters, 
have very small eccentricities in the range $e=10^{-7}-10^{-3}$ \citep[{{\it ATNF Pulsar Catalogue}},][]{mhth05}. 
Pulsar systems in globular clusters, on the other hand, often have their orbits perturbed after the recycling phase terminates because 
of their location in a dense environment \citep{rh95,hr96}.
Until the start of 2013, the only known fully recycled MSP with a high eccentricity, and located in the Galactic field, was PSR~J1903+0327 \citep[$e=0.44$,][]{crl+08}. 
This MSP has a G-type main-sequence companion star and is thought to have originated from a hierarchical triple system that ejected 
one of its members \citep{fbw+11,pvvn11,pcp12}. 

\subsection{Discovery of MSPs in eccentric orbits}
Recently, \citet{dsmb+13} presented the discovery of PSR~J2234+06 and soon 
afterwards \citet{bck+13} announced the discovery of PSR~J1946+3417. 
PSR~J2234+06 and PSR~J1946+3417 are of special interest because they resemble each other and share very unusual properties. 
Both of these Galactic field pulsars (see Table~\ref{table:MSPs}) have a spin period, $P\simeq 3\;{\rm ms}$, 
an orbital period, $P_{\rm orb}\simeq 30\;{\rm days}$ and a companion mass, $M_2 \simeq 0.24\;M_{\odot}$. 
All these values are within typical ranges expected for MSPs with He~WD companions. However, both of these MSP binaries are also eccentric, $e\simeq 0.13$,
which is unusual and unexpected from current formation theories of MSPs, as explained above. 
Therefore, it is clear that these system must have a formation history which is different from the ``normal'' MSP-WD systems in the Galactic field.

We notice that the two median expectations for the companion masses of PSR~J2234+06 and PSR~J1946+3417 are very similar to each other 
and close to the values expected from the correlation between WD mass and orbital period for post-LMXB systems \citep[e.g.][]{ts99}; 
in particular if the slight widening of the orbit from the event that imparted the eccentricity is accounted for, see Section~\ref{sec:properties}. 
This provides confidence that the current companions are indeed He~WDs which have lost their hydrogen envelopes 
via stable Roche-lobe overflow. Optical detections would confirm this. 

\subsection{A triple system formation scenario?}\label{subsec:triple}
By analogy with PSR~J1903+0327, one could advance the hypothesis that both PSRs~J2234+06 and J1946+3417 originated as
hierarchical triple systems, which evolved to produce a neutron star orbited by two F/G-type dwarfs. 
Because of the widening of the inner orbit during the subsequent neutron star accretion in the LMXB phase, 
the systems later became dynamically unstable \citep[e.g.][]{mik08} and one of the components was eventually ejected. The only difference being
that it was the donor star (the WD progenitor) in the inner binary which was ejected in the case of PSR~J1903+0327 \citep{fbw+11,pvvn11,pcp12},
whereas it would have been the outer tertiary star in the cases of PSR~J2234+06 and PSR~J1946+3417. 
In a triple system, the Kozai process \citep{koz62} may lead to large cyclic variations in the inner orbital eccentricity prior to ejection of the tertiary star \citep[e.g.,][]{ma01}.
Hence, one may expect a wide range of eccentricities of the surviving MSP-WD binaries.
Whether or not triple star evolution, or formation and ejection of a binary system from a dense cluster, is plausible for the relatively small 
eccentricities ($e\simeq 0.13$) observed in PSRs~J2234+06 and J1946+3417 (compared to $e=0.44$ for PSR~J1903+0327) requires detailed modelling beyond the scope of this Letter.

Here we advocate for another solution. In Section~\ref{sec:hypothesis} we present a hypothesis of a new direct formation channel of MSPs which can exactly 
explain both the unusual properties and the similarities of the recently discovered MSPs. In Section~\ref{sec:properties} we 
present simulations and make further falsifiable predictions about these systems which can be tested in the near future. 
In Section~\ref{sec:future} we discuss future perspectives and we summarise our conclusions in Section~\ref{sec:summary}.

\begin{table}
\center
\caption{Physical parameters of two newly discovered binary MSPs 
         \citep[data taken from][]{dsmb+13,bck+13}.}
\begin{tabular}{lcccc}
\hline Pulsar & $P$ & $P_{\rm orb}$ & $M_{\rm 2, med}$ & $e$ \\
\hline 
\noalign{\smallskip} 
	J2234+06   & 3.6~ms & 32~days & $0.23\;M_{\odot}$ & 0.13 \\
	J1946+3417 & 3.1~ms & 27~days & $0.24\;M_{\odot}$ & 0.14 \\
\noalign{\smallskip} 
\hline
\end{tabular}
\label{table:MSPs}
\end{table}


\section{Direct formation of millisecond pulsars via delayed AIC}\label{sec:hypothesis}
Besides from formation via core-collapse supernovae, it has been suggested for many years that neutron stars may also be produced from  
accretion-induced collapse (AIC) of a massive ONeMg~WD in a close binary \citep{nmsy79,tv86}. The properties of such neutron stars 
are unknown. It has been suggested that AIC events cannot produce MSPs directly since r-mode instabilities would spin-down any young, hot MSP
on a very short timescale \citep{aks99}. However, if the scenario described here is confirmed by further observations, then the role of r-mode instabilities
has to be revised. 

In the following, we rely on the results of the recent modelling by \citet{tsyl13}. 
They investigated a scenario where MSPs are produced {\em indirectly} via AIC, i.e. the AIC leaves behind a normal neutron star which is subsequently  
recycled to become an MSP, once the mass-transfer resumes after the donor star refills its Roche lobe and continues LMXB evolution until the end.
Their main result is that as a consequence of the finetuned mass-transfer rate necessary to make the WD grow in mass, the resultant MSPs created via the AIC channel
preferentially form with $10<P_{\rm orb}<60\;{\rm days}$, clustering more at $P_{\rm orb}\simeq 20-40\;{\rm days}$.
Furthermore, the modelling of these systems produced He~WD companions with masses, $M_{\rm WD}\simeq 0.24-0.31\;M_{\odot}$.
These values are interesting since they match exactly the observed values of $P_{\rm orb}$ and $M_{\rm WD}$ for the newly discovered 
MSPs in eccentric orbits (Table~\ref{table:MSPs}). 
However, in the \citet{tsyl13} scenario of indirect formation of MSPs, 
continued post-AIC mass transfer leads to highly circularised systems. Therefore, that scenario cannot explain the newly discovered MSPs with $e\sim 0.1$.

\subsection{Rotationally-delayed accretion-induced collapse (RD-AIC)}
In case a mass-gaining WD is spun~up to rapid rotation via near-Keplerian disk accretion \citep[][]{ldwh00}, 
it can avoid AIC \citep{yl04,yl05} and evolve further to super-Chandrasekhar mass values via continuous accretion
\citep[cf. fig.~7 in][for the possible growth up to $\ga 2\;M_{\odot}$]{tsyl13}.

\begin{figure}
\begin{center}
 \includegraphics[width=0.50\textwidth, angle=0]{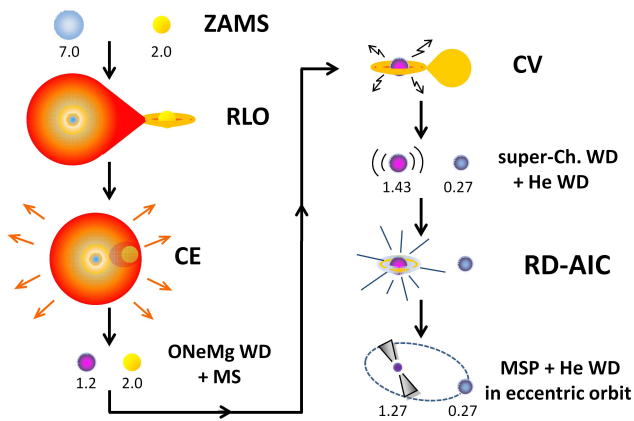}
  \caption[]{
        Illustration of the binary stellar evolution from the zero-age main
        sequence (ZAMS) to the final millisecond pulsar (MSP) stage. A primary $6-8\;M_{\odot}$ star
        evolves to initiate Roche-lobe overflow (RLO) towards the $\sim\!2\;M_{\odot}$ secondary star, 
        leading to dynamically unstable mass transfer and the formation
        of a common envelope \citep[CE;][]{ijc+13}. The envelope ejection leads to formation of an oxygen-neon-magnesium white dwarf (ONeMg~WD)
        from the naked core of the primary star (possibly after a stage of Case~BB RLO from the naked core -- not shown).
        When the secondary star evolves, it initiates RLO leading to a cataclysmic variable (CV)
	X-ray binary system.
        As a result of accretion the WD becomes a rapidly spinning super-Chandrasekhar mass WD.  
        After accretion has terminated it loses spin angular momentum and 
        eventually undergoes a rotationally-delayed accretion-induced collapse (RD-AIC)
        to directly form an MSP with a helium white dwarf (He~WD) companion in an eccentric orbit.
        Stellar masses given in units of $M_{\odot}$.
    }
\label{fig:vdh}
\end{center}
\end{figure}

Here we propose a scenario, where accretion leads to the formation of a super-Chandrasekhar mass ONeMg~WD which initially 
avoids AIC as a result of rapid rotation. Only after the accretion has terminated, and the WD loses sufficient spin angular momentum (see below), 
does it undergo AIC to {\it directly} produce an MSP. We shall refer to this event as rotationally-delayed accretion-induced collapse (RD-AIC), see Fig.~\ref{fig:vdh}.  

It is important to notice that under the new hypothesis presented here, accretion ceases
completely {\it before} AIC occurs. 
At that stage the detached system consists of two WDs: a low-mass He~WD (the remnant of the
former donor star) and an ONeMg~WD with a mass above the Chandrasekhar limit, and which later undergoes RD-AIC.
Hence, in this case there will be no re-circularisation after the AIC event.

\subsubsection{Observational and theoretical support for RD-AIC}
Observations of binaries confirm that accreting WDs rotate much faster than isolated ones \citep{sio99}; 
in one case, HD~49798/RX~J0648, there is even evidence for a WD rotating with a spin period of only $P_{\rm WD}=13.2\;{\rm s}$ \citep{mlt+11}, 
corresponding to $\sim\!50$~per~cent of its critical (break-up) rotation frequency. 
The observational evidence for such fast rotation supports the increase of the mass stability limit above 
the standard value for non-rotating WDs ($1.37\;M_{\odot}$), as required by our scenario.
An analogous idea of rotationally-delayed SNe~Ia explosions has been proposed by \citet{jus11} and \citet{dvc11} for massive CO~WDs.  

For WDs with rigid body rotation, the resulting super-Chandrasekhar masses are in the range 
$1.37-1.48\;M_{\odot}$ \citep[][and references therein]{yl04}. 
For differentially rotating WDs, the stability limit may in principle reach $\sim\!4.0\;M_{\odot}$, although   
it is quite possible that efficient transport of angular momentum by magnetic torques and/or baroclinic instabilities 
acts to ensure rigid rotation \citep{pir08}.  
On the other hand, recent observations of exceptionally luminous SNe~Ia \citep[e.g.][]{hsn+06,saa+10} suggest that their WD progenitors had a
mass of $\sim\!2.0-2.5\;M_{\odot}$.
If the critical rotation frequency is obtained during accretion then further mass accumulation is prohibited, 
unless angular momentum is transported from the WD back to the disk by viscous effects \citep{pn91,sn04}.

The final fate of super-Chandrasekhar ONeMg WDs depends on whether or not the effects of electron captures dominate over nuclear burning \citep{nmsy79,nk91}.
The onset of electron captures on $^{24}{\rm Mg}$ and $^{20}{\rm Ne}$ occurs at a density of $\rho \sim\!4\times 10^9\;{\rm g}\,{\rm cm}^{-3}$, 
whereas the density for the ignition of explosive nuclear burning (oxygen deflagration) depends on the central temperature.
Therefore, after accretion has terminated, the final fate of a super-Chandrasekhar WD depends on the competition between its cooling rate and its loss of angular momentum, 
as demonstrated in detail by \citet{yl04,yl05}. If the WD interior has crystallised by the time its spin angular momentum decreases below the critical level
(corresponding to $J^{\rm AIC}_{\rm crit}\simeq 0.4\times 10^{50}\;{\rm erg}\,{\rm s}$, for a $1.48\;M_{\odot}$ WD) it undergoes RD-AIC. 

\citet{yl04,yl05} discussed the loss of WD spin angular momentum due to gravitational wave emission caused by so-called CFS instabilities to non-axisymmetric perturbations.
In their second paper, these authors investigated 2-dimensional models and found that only r-mode instabilities \citep{and98} are relevant for accreting WDs, 
whereas bar-mode instabilities \citep{cha70,fs78} are irrelevant because the ratio of rotational to potential energy cannot reach the critical limit of $T/W=0.14$ 
(corresponding to $J=4\times10^{50}\;{\rm erg}\,{\rm s}$). 
The estimated timescale of removing (or redistributing) angular momentum has been estimated to be in the range $10^5-10^9\;{\rm yr}$, 
depending on $T/W$ and the degree of differential rotation of the WD \citep{lin99,yl04,yl05}.
However, recent work by \citet{is12} questions the efficiency of r-mode instabilities and hence they advocate for a very long 
delay timescale $>1\;{\rm Gyr}$. This would give the super-Chandrasekhar mass WD plenty of time to cool down, crystallise and undergo RD-AIC, thus favouring our scenario.

To summarize, we postulate that MSPs can be formed directly (without any need for further spin~up from a companion star) in an RD-AIC event 
that happens up to $\sim\!1\;{\rm Gyr}$ after termination of the mass-transfer phase. 

\begin{figure}
\begin{center}
\includegraphics[width=0.32\textwidth, angle=-90]{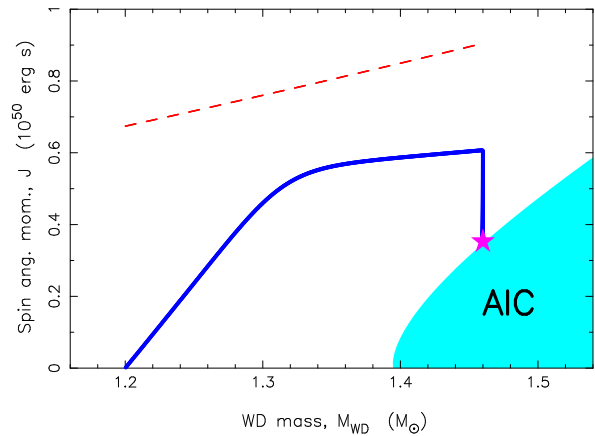}
  \caption[]{
        Schematic evolutionary track (blue solid line) in the ($M_{\rm WD},J$)--plane calculated for an accreting ONeMg~WD in a close binary system.
        After termination of the mass-transfer process, the super-Chandrasekhar mass WD is rapidly spinning, which prevents its collapse. 
        From this point, the spin evolution is solely determined by loss of angular momentum (e.g. as caused by r-mode instabilities and magnetodipole radiation).
        The light-blue hatched area marks the critical region, $J\le J_{\rm crit}^{\rm AIC}$ \citep{yl05} at which boundary the WD undergoes a RD-AIC event and produces a MSP.
        The red dashed line indicates critical break-up rotation.
    }
\label{fig:WDevol}
\end{center}
\end{figure}

In Fig.~\ref{fig:WDevol} we show an evolutionary track of a rapidly spinning WD undergoing RD-AIC
\citep[see fig.~11 in][for more detailed tracks]{yl05}.
The WD is assumed to be non-spinning initially and have a mass of $1.2\;M_{\odot}$ prior to accretion from its companion star. 
We assumed rigid rotation and efficient angular momentum accretion at the Keplerian disk value. 
The r-mode instabilities (giving rise to loss of rotational energy via gravitational waves) were calculated during accretion following
\citet{lin99}. If the timescale of loss of spin angular momentum, from the termination of the accretion phase until the WD has a spin 
angular momentum, $J<J_{\rm crit}^{\rm AIC}$, is sufficiently long ($\sim\!10^9{\rm yr}$) then the result is an AIC event \citep{yl05}.

\section{Properties of the RD-AIC events and resultant MSP-WD systems}\label{sec:properties}
The implosion of a WD with a radius of about 3000~km and an assumed surface magnetic flux density, $B\sim\!10^3\;{\rm G}$ \citep[e.g.][]{jan+07} into a neutron star with a radius of
$\sim\!10\;{\rm km}$ should produce, by conservation of magnetic flux, an MSP surface B-field of $10^3\;{\rm G} \times (3000/10)^2 \sim 10^8\;{\rm G}$.
The resultant neutron star must have a spin rate below the break-up limit and for a typical MSP spin period of a few~ms, it is expected that it must lose 
spin angular momentum during the AIC, possibly by ejection of a~few~$0.01\;M_{\odot}$ of baryonic matter in a circumstellar disk. 
According to modelling by \citet{dbo+06,kjh06,mpq09,dmq+10}, up to a few $0.01\;M_{\odot}$ of material is ejected in the AIC event, possibly leading to
synthesis of $^{56}{\rm Ni}$ in the disk which may result in a radioactively powered, short-lived SN-like transient (that peaks within $\le 1\;{\rm day}$ 
and with a bolometric luminosity $\simeq 10^{41}\;{\rm erg}\,{\rm s}^{-1}$). 

The RD-AIC hypothesis makes several very precise, easily falsifiable predictions:  
\begin{itemize}
\item As already mentioned, the He~WD companions in our RD-AIC scenario are expected to have masses in the range $M_2\simeq 0.24-0.31\;M_{\odot}$ 
(up to $0.35\;M_{\odot}$ for low-metallicity WD progenitors) and orbital periods of $10-60\;{\rm days}$.
In rare cases, we expect WD masses up to $\sim 0.41\;M_{\odot}$, if the donor star had a ZAMS mass $>2.3\;M_{\odot}$ \citep{tsyl13}. 

\item The binding energy of a neutron star can be expressed as: $E_b\simeq 0.084\,(M_{\rm NS}/M_{\odot})^2\;M_{\odot}\,c^2$ \citep{ly89},
where $M_{\rm NS}$ is its gravitational mass. 
The collapse of super-Chandrasekhar mass WDs of $1.37-1.48\;M_{\odot}$ (for rigid rotation) therefore leads to MSPs with gravitational masses of $1.22-1.31\;M_{\odot}$, if we assume
that $0.02\;M_{\odot}$ of baryonic material is lost during the AIC.

\item The sudden release of gravitational binding energy (and mass ejection into a disk) increases the orbital period and imposes an
eccentricity to the system given by \citep{bv91}:
$e=\Delta M/(M_{\rm NS}+M_2)$, if the AIC is symmetric and no kick is imparted to the newborn MSP (see below). 
Here we assume that the pre-AIC binary orbit is circular, which is a good assumption for
X-ray binaries were tidal torques circularise the system on a short timescale. For the ranges of $M_{\rm NS}$ and $M_2$ given above, this
leads to a remarkable narrow range of post-AIC eccentricities: $0.09-0.12$. (The exact values depend on the still unknown equation-of-state of neutron stars.)
This result is in excellent agreement with the systems presented in Table~\ref{table:MSPs}, cf. Section~\ref{subsec:simulations} for a discussion.

\item The momentum kick imparted to a newborn neutron star via an AIC event is expected to be small.
This follows from detailed simulations of AIC events which imply explosion energies significantly smaller than those inferred for 
standard iron-core collapse supernovae \citep{kjh06,dbo+06}, 
and also because of the small ejecta mass and the short timescale of the event (compared to the timescales of the non-radial hydrodynamic instabilities producing large kicks), 
e.g. \citet{plp+04,jan12}.
Our hypothesis therefore predicts that eccentric binary MSPs with He~WDs will have small peculiar space velocities.
\end{itemize}

\subsection{Simulations of the $(P_{\rm orb},e)$--plane}\label{subsec:simulations}
\begin{figure}
\begin{center}
 \includegraphics[width=0.34\textwidth, angle=-90]{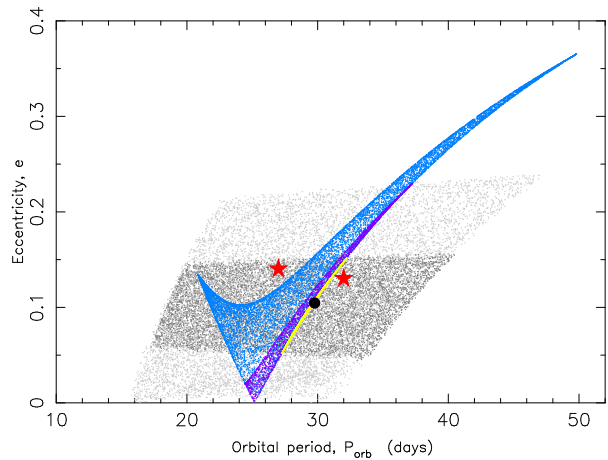}
  \caption[]{
    Distribution in the $(P_{\rm orb},e)$--plane of systems formed via the RD-AIC scenario. 
    The two red stars are the recently discovered eccentric MSPs (Table~\ref{table:MSPs}).
    The black solid circle indicates a symmetric ($w=0$) AIC from a $1.43\;M_{\odot}$ WD with a pre-AIC orbital period, $P_{\rm orb,0}=24\;{\rm days}$ 
    and a He~WD companion star of mass, $M_2=0.27\;M_{\odot}$.
    The V-shaped light-blue distribution is for the same system but applying a small kick of $w=10\;{\rm km}\,{\rm s}^{-1}$ in a random (isotropic) direction. 
    The indigo-violet and the yellow distributions superimposed are also for the same system but with $w=5$ and $2\;{\rm km}\,{\rm s}^{-1}$, respectively.
    The wide light grey distribution is for $w=5\;{\rm km}\,{\rm s}^{-1}$ and a random selection of pre-AIC systems (assuming equal probabilities),  
    with $1.37-1.48\;M_{\odot}$ WDs 
    and $P_{\rm orb,0}=15-30\;{\rm days}$ (corresponding to $M_2=0.26-0.28\;M_{\odot}$). 
    The dark grey distribution is similar but restricted to $w=2\;{\rm km}\,{\rm s}^{-1}$.
    }
\label{fig:pe}
\end{center}
\end{figure}

The spread of eccentricities and orbital periods of the resultant systems formed via RD-AIC is extremely
sensitive to any kick given to the MSP during the AIC event. 
In Fig.~\ref{fig:pe} we demonstrate this by showing a Monte Carlo simulation of the expected eccentricities
and orbital periods using the range of pre-AIC parameters given above and adding small kick velocities of $w\le 10\;{\rm km}\,{\rm s}^{-1}$.
The dynamical effects were calculated following the formulae of \cite{hil83}.
The properties of systems undergoing RD-AIC events are seen to be surprisingly similar to the characteristics of the recently discovered MSPs in eccentric orbits (Table~\ref{table:MSPs}).

\section{Future perspectives and tests}\label{sec:future}
If the WD companions happen to be bright, then a study of their spectral lines will yield the mass ratio of the binary components, $q$.
Furthermore, given the eccentric orbits of these MSPs, we will certainly be able to measure the rate of advance of periastron ($\dot{\omega}$) for these systems.
If the radius of the companion is small compared to the size of the orbit (which is the case for a WD), then $\dot{\omega}$ is solely due to the effects of general relativity and
can be used to estimate the total mass of the system \citep{wt81}. The combination of $\dot{\omega}$ and $q$ would be enough to determine the masses of the components. 
Another possible solution is the measurement of the Shapiro delay for these systems. Even a relatively low-precision measurement of $h_3$ \citep{fw10} can,
when combined with the measurement of $\dot{\omega}$, yield very precise component masses, as in the cases of PSR~J1903+0327 \citep{fbw+11} and PSR~J1807$-$2500B \citep{lfrj12}.
These mass measurements are very important for testing the RD-AIC hypothesis, which predicts MSP masses between $1.22-1.31\;M_{\odot}$.  Measuring a higher MSP mass would, if not falsifying our hypothesis, require differential rotation
of the progenitor WD, which may be a problem with respect to the need of a long delay timescale \citep{is12}. 

The unusual MSPs discussed in this Letter were discovered in recent pulsar surveys \citep[e.g.][]{cfl+06,dsmb+13,bck+13} with high time and frequency resolution that have greatly increased the
number of known MSPs, revealing new rare pulsar populations. If on-going and future surveys detect many 
eccentric MSPs with WD companions with $e\sim 0.1$ (and $P_{\rm orb}=10-60\;{\rm days}$), 
this would not only support our RD-AIC hypothesis; it would also imply that AIC events do not produce kicks (or at least $w\le 5\;{\rm km}\;{\rm s}^{-1}$, cf. Fig.~\ref{fig:pe}) 
and that WDs rotate rigidly. Furthermore, it would imply that r-mode instabilities do not necessarily slow down young, hot MSPs, as previously suggested \citep{aks99}.

Note, there may also be eccentric MSPs with WDs formed via the triple scenario outlined in Section~\ref{subsec:triple},  
which will have a much wider distribution in the $(P_{\rm orb},e)$--plane 
and possibly more massive companions.
Detection of an MSP with a main-sequence companion and $e \sim 0.1$ would support a triple star scenario for the formation of MSPs with WDs and $e\sim 0.1$, 
and thus significantly weaken the need for our RD-AIC hypothesis.

Population synthesis investigations of MSP formation via AIC have been performed by \citet{htw+10} and \citet{clxl11}.
The former study concluded that, in general, the AIC channel to MSP formation is important. The latter study investigated direct MSP formation via AIC
and concluded that the probability of forming eccentric MSPs can be ruled out (Even using high kicks they could not produce
eccentric MSPs with $P_{\rm orb}\ge 20\;{\rm days}$), in contradiction with the new discoveries, cf. Table~\ref{table:MSPs}.
We recommend new population synthesis modelling using our RD-AIC scenario in order to
probe more carefully the expected number of such eccentric MSP systems to be detected, and for the statistics of their resulting parameter space. 
Ideally, the triple system scenario should be modelled for comparison as well.

Finally, it should be investigated under which circumstances a binary evolves via RD-AIC or follows the \citet{tsyl13} path. The latter was calculated using a point mass accreting WD
which did not allow for detailed spin angular momentum modelling.
For the resulting MSPs with He~WD companions, the values of $P_{\rm orb}$ and $M_2$ are expected to be roughly similar. The RD-AIC scenario, however,
produces eccentric systems.

\section{Summary}\label{sec:summary}
The common scenario for the formation of MSPs via recycling in LMXBs is well established with plenty of observational evidence, as discussed in Section~\ref{sec:intro}. 
The RD-AIC hypothesis presented in this Letter provides an additional formation channel of MSPs that makes very specific predictions about future discoveries
and the existence of a separate population of eccentric MSPs.
If this hypothesis is confirmed by future observations, it would also have interesting consequences for better understanding 
the direct AIC channel to produce MSPs, i.e. with respect to WD progenitor masses, (absence of) momentum kicks in AIC, and possibly even constraining 
neutron star equations-of-state given that the post-AIC eccentricities depend on the released gravitational binding energy.

\section*{Acknowledgements}
We thank the anonymous referee for very constructive comments that improved this manuscript and Sung-Chul Yoon for discussions.
P.F. gratefully acknowledges the financial support by the European Research Council for the ERC Starting Grant BEACON under contract no. 279702.
T.M.T. gratefully acknowledges financial support and hospitality at both the
Argelander-Insitut f\"ur Astronomie, Universit\"at Bonn and the Max-Planck-Institut f\"ur Radioastronomie.



\footnotesize{

}
%


\end{document}